\newif\ifacm
\newif\ifspacestretch
\documentclass[sigplan,nonacm]{acmart}
\ifacm
\else
\renewcommand\footnotetextcopyrightpermission[1]{} % removes footnote with conference information in first column
\fi

\usepackage{todo}
\usepackage{graphicx}
\usepackage{amsmath}
\usepackage{amsthm}
\usepackage{amsfonts}
\usepackage{amsthm}
\usepackage{xcolor}
\usepackage{verbatim}
\usepackage{listings}
\usepackage{tikz}
\usepackage{hyperref}
\usepackage{booktabs}
\usepackage{multirow}
\usepackage{enumitem}
\usepackage{placeins}
\usepackage{xfrac}
\usetikzlibrary{positioning}
\usetikzlibrary{fit}
\newfloat{lstfloat}{htbp}{lop}
\floatname{lstfloat}{Listing}

\setlist[enumerate]{leftmargin=4ex}
\setlist[itemize]{leftmargin=4ex}

\newcommand{\articletitle}{CesASMe and Staticdeps: static detection of memory-carried dependencies for code analyzers}

\author{Théophile Bastian}
\affiliation{\institution{Univ. Grenoble Alpes, Inria, CNRS, Grenoble INP, LIG, 38000 Grenoble, France} \country{}}
\email{theophile.bastian@inria.fr}

\author{Hugo Pompougnac}
\affiliation{\institution{Univ. Grenoble Alpes, Inria, CNRS, Grenoble INP, LIG, 38000 Grenoble, France} \country{}}
\email{hugo.pompougnac@inria.fr}

\author{Alban Dutilleul}
\affiliation{\institution{ENS Rennes} \country{France}}
\email{alban.dutilleul@ens-rennes.fr}

\author{Fabrice Rastello}
\affiliation{\institution{Univ. Grenoble Alpes, Inria, CNRS, Grenoble INP, LIG, 38000 Grenoble, France} \country{}}
\email{fabrice.rastello@inria.fr}
\title{\articletitle}
\date{}

\newcommand{\iaca}{\texttt{IACA}}
\newcommand{\gus}{\texttt{Gus}}
\newcommand{\llvmmca}{\texttt{llvm-mca}}
\newcommand{\uica}{\texttt{uiCA}}
\newcommand{\ithemal}{\texttt{Ithemal}}
\newcommand{\osaca}{\texttt{Osaca}}
\newcommand{\bhive}{\texttt{BHive}}
\newcommand{\anica}{\texttt{AnICA}}
\newcommand{\valgrind}{\texttt{valgrind}}
\newcommand{\depsim}{\texttt{depsim}}
\newcommand{\staticdeps}{\texttt{staticdeps}}
\newcommand{\Staticdeps}{\texttt{Staticdeps}}
\newcommand{\vex}{\texttt{VEX}}
\newcommand{\uicadeps}{\uica{}~+~\staticdeps{}}

\newcommand{\perf}{\texttt{perf}}

\newcommand{\qemu}{\texttt{QEmu}}

\newcommand{\cesasme}{\texttt{CesASMe}}

\newcommand{\eg}{\textit{eg.}}
\newcommand{\ie}{\textit{ie.}}
\newcommand{\wrt}{\textit{wrt.}}

\newcommand{\reg}[1]{\texttt{\%#1}}

\newcommand{\uop}{$\mu$OP}
\newcommand{\uops}{\uop{}s}
\newcommand{\kerK}{\mathcal{K}}
\newcommand{\calR}{\mathcal{R}}
\newcommand{\calRbar}{\overline{\calR}}
\newcommand{\cov}{\operatorname{cov}}

\newcommand{\card}[1]{\left| #1 \right|}

\definecolor{lessimportantgrey}{HTML}{707070}

\newcommand{\figref}[1]{[\ref{#1}]}

\newtheorem{definition}{Definition}
\newtheorem{postulate}{Postulate}
\newtheorem{lemma}{Lemma}
\newtheorem{thm}{Theorem}
\newcommand{\distance}[2]{\textrm{distance}\left(#1,#2\right)}

\newcommand{\lstxasm}[1]{\lstinline[language=={[x86masm]Assembler}]$#1$}

\ifspacestretch
\else
\fi

\begin{document}
\begin{abstract}
    A variety of code analyzers, such as \iaca, \uica, \llvmmca{} or
    \ithemal{}, strive to statically predict the throughput of a computation
    kernel.  Each analyzer is based on its own simplified CPU model
    reasoning at the scale of a basic block.
    Facing this diversity, evaluating their strengths and
    weaknesses is important to guide both their usage and their enhancement.

    We present \cesasme{}, a fully-tooled solution to evaluate code analyzers
    on C-level benchmarks composed of a benchmark derivation procedure that
    feeds an evaluation harness. We conclude that memory-carried data
    dependencies are a major source of imprecision for these tools. We tackle
    this issue with \staticdeps{}, a static analyzer extracting
    memory-carried data dependencies, including across loop iterations, from an
    assembly basic block. We integrate its output to \uica{}, a
    state-of-the-art code analyzer, to evaluate \staticdeps{}' impact on a code
    analyzer's precision through \cesasme{}.
\end{abstract}

\maketitle

\section{Introduction}\label{sec:intro}

At a time when software is expected to perform more computations, faster and in
more constrained environments, tools that statically predict the resources
they consume are very useful to guide
optimizations. This need is reflected in the diversity of binary or assembly
code analyzers that appeared following the deprecation of \iaca~\cite{iaca}, which Intel has
maintained through 2019. Whether it is \llvmmca{}~\cite{llvm-mca},
\uica{}~\cite{uica}, \ithemal~\cite{ithemal} or \gus~\cite{phd:gruber}, all
these tools strive to produce various
performance metrics, including the number of CPU cycles a computation kernel will take
---~which roughly translates to execution time.
In addition to raw measurements relying on hardware counters, these model-based
analyses provide
higher-level and refined data, to expose the bottlenecks and guide the
optimization of a given code. This feedback is useful to experts optimizing
computation kernels, including scientific simulations and deep-learning
kernels.

An exact throughput prediction would require a cycle-accurate simulator of the
processor such as gem5~\cite{gem5}, but would also require 
microarchitectural data that is most often not publicly available, 
and would be prohibitively slow in any case. These tools thus each 
solve in their own way the challenge of modeling complex CPUs while remaining
simple enough to yield a prediction in a reasonable time, ending up with
different models.

In particular, previous works (\eg{} \anica{}~\cite{anica}) have
exposed significant discrepancies in the reported predictions
for programs with memory-carried dependencies.
This is a clue of a more general limitation of current code
analyzers, which leads to the hypothesis that
the underlying performance models have trouble with
memory dependencies modelization.
One way to verify this is to provide an alternative analysis
method and be able to show that it produces more accurate predictions.
Such an approach will also need an evaluation framework sound enough
to trustworthy compare one method with another.

\subsection{Contributions}\label{ssec:contribs}

We present a new method to construct sound baselines against which
to compare the predictions of code analyzers, in order to evaluate
the performance models underlying the latter.
Indeed, existing approaches lack such a baseline
and often require \textit{ad hoc}, handmade supplementary analysis.
Some evaluate the results of a code analyzer by comparing it to those of
another code analyzer (\eg{} \anica), while
others perform actual measurements, but on
---~possibly randomly-generated~--- basic blocks isolated from their context,
\eg{} using \bhive{}~\cite{bhive}.

By comparison, our approach, which is materialized in a tool called
\cesasme{}, consists in generating various
(L1-resident, code analyzers-compliant) \textit{microbenchmarks}
from an existing benchmark suite, and then using them without isolating
their constituent basic blocks to evaluate
code analyzers. To do so, \cesasme{} applies static analyses on their
constituent basic blocks. At the same time, it
executes the microbenchmarks and performs hardware counters-based measurements.
It finally aggregates measurements and analyses in such a way as to make them
commensurable.

We also provide a new heuristic called
\staticdeps{} to extract memory-carried dependencies.
We use this heuristic with \cesasme{} to test the hypothesis mentioned above
that current code analyzers struggle to model the latter.
We show that this approach
significantly improves the accuracy of the code analyzer
\uica{}\cite{uica}, which is tantamount to showing that
dependencies constitute a significant limitation of its underlying model.
We also show that the resulting extended analyzer fully outperforms the state
of the art.

\subsection{Related works}

The static throughput analyzers studied rely on a variety of models.
\iaca~\cite{iaca}, developed by Intel and now deprecated, is closed-source and
relies on Intel's expertise on their own processors.
The LLVM compiling ecosystem, to guide optimization passes, maintains models of many
architectures. These models are used in the LLVM Machine Code Analyzer,
\llvmmca~\cite{llvm-mca}, to statically evaluate the performance of a segment
of assembly.
Independently, Abel and Reineke used an automated microbenchmark generation
approach to generate port mappings of many architectures in
\texttt{uops.info}~\cite{nanobench, uopsinfo} to model processors' backends.
This work was continued with \uica~\cite{uica}, extending this model with an
extensive frontend description.
Following a completely different approach, \ithemal~\cite{ithemal} uses a deep
neural network to predict basic blocks throughput. To obtain enough data to
train its model, the authors also developed \bhive~\cite{bhive}, a profiling
tool working on isolated basic blocks.

Another static tool, \osaca~\cite{osaca2}, provides lower- and
upper-bounds to the execution time of a basic block. As this kind of
information cannot be fairly compared with tools yielding an exact throughput
prediction, we exclude it from our scope.

All these tools statically predict the number of cycles taken by a piece of
assembly or binary that is assumed to be the body of an infinite ---~or
sufficiently large~--- loop in steady state, all its data being L1-resident. As
discussed by \uica's authors~\cite{uica}, hypotheses can subtly vary between
analyzers; \eg{} by assuming that the loop is either unrolled or has control
instructions, with non-negligible impact. Some of the tools, \eg{} \ithemal,
necessarily work on a single basic block, while some others, \eg{} \iaca, work
on a section of code delimited by markers. However, even in the second case,
the code is assumed to be \emph{straight-line code}: branch instructions, if
any, are assumed not taken.

Throughput prediction tools, however, are not all static.
\gus~\cite{phd:gruber} dynamically predicts the throughput of a whole program
region, instrumenting it to retrieve the exact events occurring through its
execution. This way, \gus{} can detect bottlenecks more finely through
sensitivity analysis, at the cost of a significantly longer run time.

The \bhive{} profiler~\cite{bhive} takes another approach by performing
basic block throughput \textit{measurement} instead of analysis:
by mapping memory at any address accessed by a basic
block, it can effectively run and measure arbitrary code without context, often
---~but not always, as we discuss later~--- yielding good results as one would expect
from an actual measurement.

The \anica{} framework~\cite{anica} also attempts to evaluate throughput
predictors by finding examples on which they are inaccurate. \anica{} starts
with randomly generated assembly snippets, and refines them through a process
derived from abstract interpretation to reach general categories of problems.

\section{Benchmarking harness}\label{sec:bench_harness}

The obvious solution to evaluate predictions is to compare them to an
actual measure. However, as these tools reason at the basic block level, this
is not as trivially defined as it would seem,
especially when it comes to showing the impact of dependencies.
As detailed later in
Section~\ref{ssec:bhive_errors}, \emph{without proper context, a basic
block's throughput is not well-defined}.

To recover the context of each basic block, we reason instead at the scale of
a C source code. This
makes the measures unambiguous: one can use hardware counters to measure the
elapsed cycles during a loop nest. This requires a suite of benchmarks, in C,
that both is representative of the domain studied, and wide enough to have a
good coverage of the domain. However, this is not in itself sufficient to
evaluate static tools:
the number of cycles reported by the counters can be the result of an
arbitrarily large number of loop turns, with conditional branches taken
or not.
This number hardly compares to \llvmmca{}, \iaca{}, \ithemal{}, and \uica{}
basic block-level predictions seen above that only predict throughput of
individual basic block.

\begin{figure*}[ht!]
    \definecolor{col_bench_gen}{HTML}{5a7eff}
    \definecolor{col_bench_gen_bg}{HTML}{dbeeff}
    \definecolor{col_bench_harness}{HTML}{ffa673}
    \definecolor{col_results}{HTML}{000000}
\centerline{
  \begin{tikzpicture}[
      hiddennode/.style={rectangle,draw=white, very thick, minimum size=5mm, align=center, font=\footnotesize},
      normnode/.style={rectangle,draw=black, very thick, minimum size=5mm, align=center, font=\footnotesize},
  resultnode/.style={rectangle,draw=col_results, fill=black!2, very thick, minimum size=5mm, align=center, font=\footnotesize},
  bluenode/.style={rectangle, draw=col_bench_gen, fill=col_bench_gen_bg, very thick, minimum height=5mm, minimum width=4cm, align=center, font=\footnotesize},
  rednode/.style={rectangle, draw=col_bench_harness, fill=orange!5, very thick, minimum size=5mm, align=center, font=\footnotesize},
  bencher/.style={rednode, minimum width=2.5cm, minimum height=5mm},
  genarrow/.style={draw=col_bench_gen},
  harnarrow/.style={draw=col_bench_harness},
  ]
  \centering
  %Nodes
  \node[bluenode]  (bench)                       {Benchmark suite};
  \node[bluenode]  (pocc)     [below=of bench] {Loop nest optimizers};
  \node[bluenode]  (kernel)   [below=of pocc]    {Microkernel extraction};
  \node[bluenode]  (gcc)      [below=of kernel]  {Compilations};
  \node[rednode]   (gdb)      [right=0.1\textwidth of gcc] {Basic block \\extraction \figref{ssec:bb_extr}};
  \node[bencher]   (ithemal)  [right=4cm of gdb]   {Ithemal};
  \node[bencher]   (iaca)     [above=0.5em of ithemal]    {IACA};
  \node[bencher]   (uica)     [above=0.5em of iaca]    {uiCA};
  \node[bencher]   (llvm)     [above=0.5em of uica]     {llvm-mca};
  \node[bencher]   (bhive)    [above=0.5em of llvm]     {BHive (measure)};
  \node[rednode]   (ppapi)    [left=1cm of bhive]   {perf (measure)};
  \node[rednode]   (gus)      [below=0.5em of ppapi] {Gus};
  %% \node[rednode]   (uica)     [below=of gdb]     {uiCA};
  \node[rednode]   (lifting)  [right=of bhive]   {
      Prediction lifting\\\figref{ssec:harness_lifting}};
  \node[
    draw=black,
    very thick,
    dotted,
    fit=(ppapi) (gus) (bhive) (llvm) (uica) (iaca) (ithemal)
  ] (comps) {};
  \node (throughput_label)  [above=0.2em of comps,align=center] {
          \footnotesize Throughput predictions \\\footnotesize \& measures
      \figref{ssec:throughput_pred_meas}};
  \node[draw=black,
    very thick,
    dotted,
    %% label={below:\footnotesize Variations},
    label={[above,xshift=1.5cm]\footnotesize Transformations \figref{sec:bench_gen}},
    fit=(pocc) (kernel) (gcc)
  ] (vars) {};
\node[resultnode]  (bench2) [below=of lifting] {Evaluation metrics \\ for
        code analyzers};

  % Key
  \node[]  (keyblue1)     [below left=0.7cm and 0cm of vars]   {};
  \node[hiddennode]  (keyblue2)     [right=0.5cm of keyblue1]   {Generating microbenchmarks};
  \node[]  (keyred1)     [right=0.6cm of keyblue2]   {};
  \node[hiddennode]  (keyred2)     [right=0.5cm of keyred1] {Benchmarking harness};
  \node[]  (keyresult1)     [right=0.6cm of keyred2]   {};
  \node[hiddennode]  (keyresult2)     [right=0.5cm of keyresult1]
      {Results analysis};

  %Lines
    \draw[-, very thick, harnarrow] (keyred1.east)     -- (keyred2.west);
    \draw[-, very thick, genarrow]  (keyblue1.east)     -- (keyblue2.west);
    \draw[-, very thick]  (keyresult1.east)     -- (keyresult2.west);
  \draw[->, very thick, genarrow]   (bench.south)   -- (pocc.north);
  \draw[->, very thick, genarrow]   (pocc.south)    -- (kernel.north);
  \draw[->, very thick, genarrow]   (kernel.south)  -- (gcc.north);
  \draw[->, very thick, genarrow]   (gcc.east)     -- (gdb.west);
  \draw[->, very thick, genarrow]   (gcc.east)     -- (ppapi.west);
  \draw[->, very thick, genarrow]   (gcc.east)     -- (gus.west);
  \draw[->, very thick, harnarrow]  (gdb.east)     -- (uica.west);
  \draw[->, very thick, harnarrow]  (gdb.east)     -- (iaca.west);
  \draw[->, very thick, harnarrow]  (gdb.east)     -- (ithemal.west);
  \draw[->, very thick, harnarrow]  (gdb.east)     -- (bhive.west);
  \draw[->, very thick, harnarrow]  (gdb.east)     -- (llvm.west);
  \draw[->, very thick, harnarrow]  (comps.east|-lifting)   -- (lifting.west);
  \draw[->, very thick]            (lifting.south)   -- (bench2.north);
  \end{tikzpicture}
}
\vspace{-1em}
\caption{Our analysis and measurement environment.\label{fig:contrib}}
\vspace{-1em}
\end{figure*}
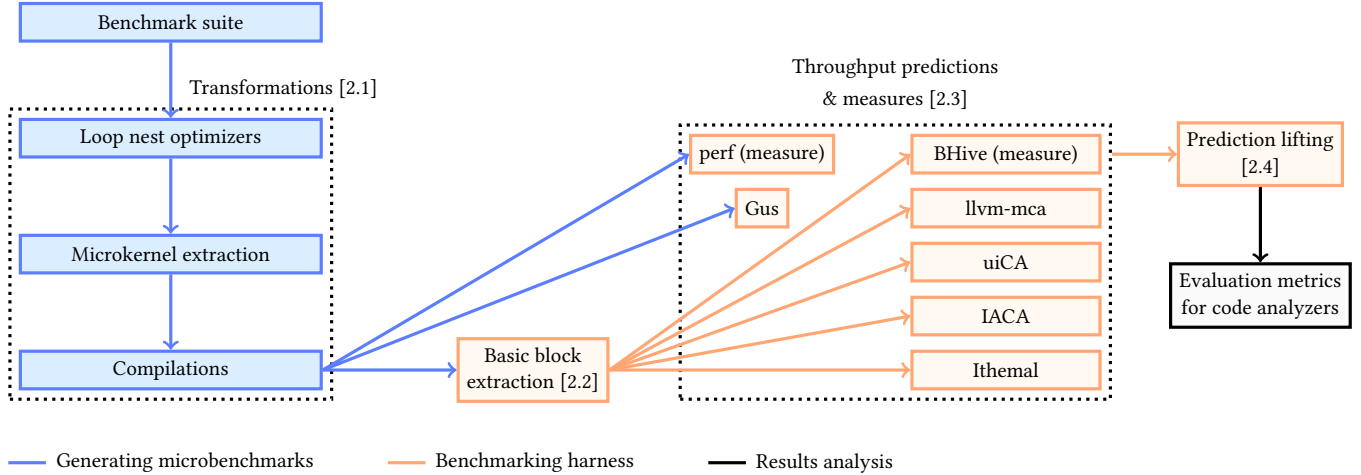

\paragraph{Predictions lifting.}

We developed a \textit{basic block occurences counter} to
count the number of occurrences of a basic
block during kernel execution. Then, we use this number of occurrences to
weight basic block throughput predictions. Finally,
we lift code analyzers' results by summing the weighted basic block
predictions:
\begin{align}\label{eqn:lifted_pred}
\text{lifted\_pred}(\mathcal{K}) =
    \sum_{b \in \operatorname{BBs}(\mathcal{K})}
    \operatorname{occurences}(b) \times \operatorname{pred}(b)
\end{align}

\paragraph{Architecture of our benchmarking harness.}
Our benchmarking framework works in four successive stages, whose
a high-level view is shown in Figure~\ref{fig:contrib}.
It first generates relevant \textit{microbenchmarks}
(\ie{} L1-resident benchmarks) by applying a series of transformations to
an existing benchmark suite which is, in our case, the HPC-native
Polybench~\cite{polybench} suite.
It then
extracts the basic blocks constituting a computation kernel, and instruments it
to retrieve their respective occurrences in the original context. It runs
all the studied tools on each basic block, while also running measures on the
whole computation kernel. Finally, the block-level results are lifted to
kernel-level results thanks to the occurrences previously measured.

\subsection{Generating microbenchmarks}\label{sec:bench_gen}

The first stage of our benchmarking harness consists in generating
a set of \textit{microbenchmarks}.
We extend \anica{}'s random-based approach mentioned above in order to
produce microbenchmarks diverse enough, but related to real-life kernels.
Thus, our approach consists in applying several \textit{transformations}
to an existing benchmark suite:
\begin{enumerate}
  \item We produce several versions of each benchmark using loop nest optimizers
    (Pluto\cite{pluto} and PoCC~\cite{pocc}). The latter gives access to a wide
    variety of transformations:
    register tiling, tiling, skewing, vectorization/simdization,
    loop unrolling, loop permutation, loop fusion and so on.
  \item We instrument the resulting code in order to extract
    the loop which has the greatest impact on
    performance while being L1-resident. This \textit{microkernel} is
    isolated from the initial benchmark, and incorporated into a parameterized
    template where requested behaviors, as the
    number of successive measurements, are set. We call this refined benchmark
    a \textit{microbenchmark}.
  \item Given a microbenchmark, we produce several binaries
    by applying different compilation strategies
    (enabling/disabling auto-vectorization, extended instruction
    sets, \textit{etc}.).
\end{enumerate}

This pipeline allows us to produce a diversified yet domain-specific range
of microbenchmarks.

\subsection{Basic block extraction}\label{ssec:bb_extr}

Given a compiled microbenchmark, we use
the Capstone disassembler~\cite{tool:capstone}, and split the assembly
code at each control flow instruction (jump, call, return, \ldots) and each
jump site.

To accurately obtain the occurrences of each basic block in the whole kernel's
computation,
we then instrument it with \texttt{gdb} by placing a break
point at each basic block's first instruction in order to count the occurrences
of each basic block between two calls to the \perf{} counters\footnote{We
assume the program under analysis to be deterministic.}.

\subsection{Throughput predictions and measures}\label{ssec:throughput_pred_meas}

The harness leverages a variety of tools: actual CPU measurement; the \bhive{}
basic block profiler~\cite{bhive}; \llvmmca~\cite{llvm-mca}, \uica~\cite{uica}
and \iaca~\cite{iaca}, which leverage microarchitectural
models to predict a block's throughput; \ithemal~\cite{ithemal}, a machine
learning model; and \gus~\cite{phd:gruber}, a dynamic analyzer based on \qemu{}
that works at the whole binary level.

The execution time of the full kernel is measured using Linux
\perf~\cite{tool:perf} CPU counters around the full computation kernel. The
measure is repeated four times and the smallest is kept; as the caches are not
flushed, this ensures that the cache is warm and compensates for context
switching or other measurement artifacts. \gus{} instruments the whole function
body. The other tools work at basic block level; these are run on
each basic block of each benchmark.

We emphasize the importance, throughout the whole evaluation chain, to keep the
exact same assembled binary. Indeed, recompiling the kernel from source
\emph{cannot} be assumed to produce the same assembly kernel. This is even more
important in the presence of slight changes: for instance, inserting \iaca{}
markers at the C-level ---~as is intended~--- around the kernel \emph{might}
change the compiled kernel, if only for alignment reasons. Furthermore, those
markers prevent a binary from being run by overwriting registers with arbitrary
values. This forces a user to run and measure a version which is different from
the analyzed one. In our harness, we circumvent this issue by adding markers
directly at the assembly (already compiled) basic block level.  Our
\texttt{gdb} instrumentation procedure also respects this principle of
single-compilation. As \qemu{} breaks the \perf{} interface, we have to run
\gus{} with a preloaded stub shared library to be able to instrument binaries
containing calls to \perf{}.

\subsection{Prediction lifting and filtering}\label{ssec:harness_lifting}

We finally lift single basic block predictions to a whole-kernel cycle
prediction by summing the block-level results, weighted by the occurrences of
the basic block in the original context (\autoref{eqn:lifted_pred} above). If an analyzer fails
on one of the basic blocks of a benchmark, the whole benchmark is discarded for
this analyzer.

%% In the presence of complex control flow, \eg{} with conditionals inside loops,
%% our approach based on basic block occurrences is arguably less precise than an
%% approach based on paths occurrences, as we have less information available
%% ---~for instance, whether a branch is taken with a regular pattern, whether we
%% have constraints on register values, etc. We however chose this block-based
%% approach, as most throughput prediction tools work a basic block-level, and are
%% thus readily available and can be directly plugged into our harness.\qtodo{À
%% garder ou pas ?}

Finally, we control the proportion of cache misses in the program's execution
using \texttt{Cachegrind}~\cite{valgrind} and \gus; programs that have more
than 15\,\% of cache misses on a warm cache are not considered L1-resident and
are discarded.

\subsection{Soundness of \cesasme{}'s methodology}

Given a time prediction
$C_\text{pred}$ (in cycles) and a baseline $C_\text{baseline}$, we define its
relative error as
\begin{align}
    \operatorname{err} = \frac{\left| C_\text{pred} - C_\text{baseline}
    \right|}{C_\text{baseline}}\label{eqn:rel_error}
\end{align}
We assess the commensurability of the whole benchmark, measured with \perf{}, to
lifted block-based results by measuring the statistical distribution of the
relative error of two series: the predictions made by \bhive, and the series of
the best block-based prediction for each benchmark.

We single out \bhive{} as it is the only tool able to \textit{measure}
---~instead of predicting~--- an isolated basic block's timing. 
Since \bhive{} is based on measures ---~instead of predictions~---
through hardware counters, an excellent accuracy is expected.
This, however, is
not sufficient: as discussed later in Section~\ref{ssec:bhive_errors}, \bhive{}
is not able to yield a result for about $40\,\%$ of the benchmarks, and is
subject to large errors in some cases.

\begin{figure}
    \centering
    \includegraphics[width=\linewidth]{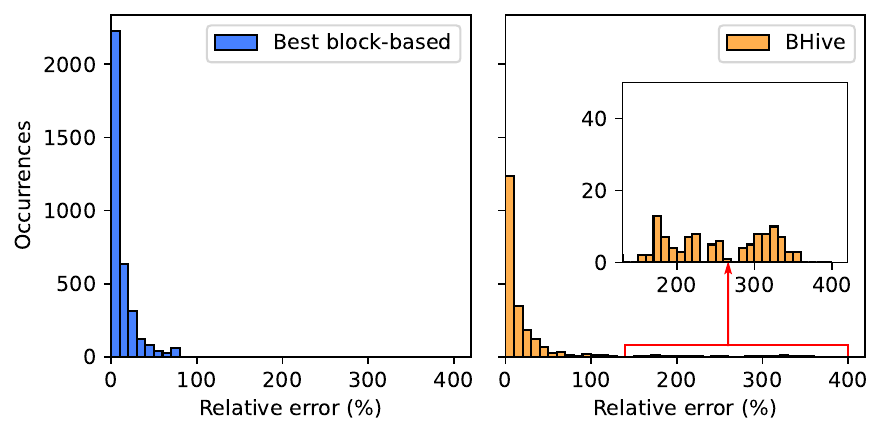}
    \vspace{-0.8cm}
    \caption{Relative error distribution \wrt{} \perf}\label{fig:exp_comparability}
\end{figure}

The result of this analysis is presented in Table~\ref{table:exp_comparability}
and in Figure~\ref{fig:exp_comparability}. The results are in a range
compatible with common results of the field, as seen \eg{} in~\cite{uica}
reporting Mean Absolute Percentage Error (MAPE, corresponding to the
``Average'' row) of about 10-15\,\% in many cases. While lifted \bhive's
average error is driven high by large errors on certain benchmarks,
investigated later in this article, its median error is still comparable to the
errors of state-of-the-art tools. From this, we conclude that lifted cycle
measures and predictions are consistent with whole-benchmark measures; and
consequently, lifted predictions can reasonably be compared to one another.

\begin{table}
    \centering
    \begin{tabular}{l r r}
        \toprule
        & \textbf{Best block-based} & \textbf{BHive} \\
        \midrule
        Datapoints & 3500 & 2198 \\
        Errors & 0 & 1302 \\
         & (0\,\%) & (37.20\,\%) \\
        Average (\%) & 11.60 & 27.95 \\
        Median (\%) & 5.81 & 7.78 \\
        Q1 (\%) & 1.99 & 3.01 \\
        Q3 (\%) & 15.41 & 23.01 \\
        \bottomrule
    \end{tabular}
    \caption{Relative error statistics \wrt{} \perf}\label{table:exp_comparability}
    \vspace{-0.8cm}
\end{table}

\subsection{Understanding \bhive's results}\label{ssec:bhive_errors}

The error distribution of \bhive{} against \perf{}, plotted right in
Figure~\ref{fig:exp_comparability}, puts forward irregularities in \bhive's
results. Its lack of
support for control flow instructions can be held accountable for a portion of
this accuracy drop; our lifting method, based on block occurrences instead of
paths, can explain another portion. We also find that \bhive{} fails to produce
a result in about 40\,\% of the kernels explored ---~which means that, for those
cases, \bhive{} failed to produce a result on at least one of the constituent
basic blocks. In fact, this is due to the need to reconstruct the context of each
basic block \textit{ex nihilo}.

The basis of \bhive's method is to run the code to be measured, unrolled a
number of times depending on the code size, with all memory pages but the
code unmapped. As the code tries to access memory, it will raise segfaults,
caught by \bhive's harness, which allocates a single shared-memory page, filled
with a repeated constant, that it will map wherever segfaults occur before
restarting the program.
\bhive{} mainly fails on bad code behaviour (\eg{} control flow
not reaching the exit point of the measure if a bad jump is inserted), too many
segfaults to be handled, or a segfault that occurs even after mapping a page at
the problematic address.

The registers are also initialized, at the beginning of the measurement, to the
fixed constant \texttt{0x2324000}. We show through two examples that this
initial value can be of crucial importance.

The following experiments are executed on an Intel(R) Xeon(R) Gold 6230R CPU
(Cascade Lake), with hyperthreading disabled.

\paragraph{Imprecise analysis} Consider this x86-64 kernel:

\begin{minipage}{0.95\linewidth}
\begin{lstlisting}[language={[x86masm]Assembler}]
    vmulsd (%rax), %xmm3, %xmm0
    vmovsd %xmm0, (%r10)
\end{lstlisting}
\end{minipage}

When executed with all the general purpose registers initialized to the default
constant, \bhive{} reports 9 cycles per iteration, since \reg{rax} and
\reg{r10} hold the same value, inducing a read-after-write dependency between
the two instructions. If, however, \bhive{} is tweaked to initialize \reg{r10}
to a value that aliases (\wrt{} physical addresses) with the value in
\reg{rax}, \eg{} between \texttt{0x10000} and \texttt{0x10007} (inclusive), it
reports 19 cycles per iteration instead; while a value between \texttt{0x10008}
and \texttt{0x1009f} (inclusive) yields the expected 1 cycle ---~except for
values in \texttt{0x10039}-\texttt{0x1003f} and
\texttt{0x10079}-\texttt{0x1007f}, yielding 2 cycles as the store crosses a
cache line boundary.

In the same way, the value used to initialize the shared memory page can
influence the results whenever it gets loaded into registers.

\paragraph{Failed analysis} Some memory accesses will always result in an
error; for instance, it is impossible (by default) to \texttt{mmap} at an address lower
than \texttt{0x10000}. Thus,
with equal initial values for all registers, the following kernel would fail,
since the second operation attempts to load at address 0:

\begin{minipage}{0.95\linewidth}
\begin{lstlisting}[language={[x86masm]Assembler}]
    subq %r11, %r10
    movq (%r10), %rax
\end{lstlisting}
\end{minipage}

Such errors can occur in more circumvoluted ways. The following x86-64 kernel,
for instance, is extracted from a version of the \texttt{durbin}
kernel:

\begin{minipage}{0.95\linewidth}
\begin{lstlisting}[language={[x86masm]Assembler}]
    vmovsd 0x10(%r8, %rcx), %xmm6
    subl %eax, %esi
    movslq %esi, %rsi
    vfmadd231sd -8(%r9, %rsi, 8), \
        %xmm6, %xmm0
\end{lstlisting}
\end{minipage}

Here, \bhive{} fails to measure the kernel when run with
registers initialized to the default constant at the 2\textsuperscript{nd}
occurrence of the unrolled loop body, failing to recover from an error at the
\texttt{vfmadd231sd} instruction with the \texttt{mmap} strategy. Indeed, after
the first iteration the value in \reg{rsi} becomes null, then negative at the
second iteration; thus, the second occurrence of the last instruction fetches
at address \texttt{0xfffffffff0a03ff8}, which is in kernel space. This
microkernel can be benchmarked with BHive \eg{} by initializing \reg{rax} to 1.
Furthermore, on most 64-bits CPUs, addresses are only valid if they are
representable on 48 bits (correctly sign-extended to fill 64
bits)~\cite{ref:intel64_software_dev_reference_vol1,ref:amd64_architecture_dev_reference_vol2},
called \emph{canonical form}, causing further failures in \bhive. Other kernels still fail with
accesses relative to the instruction pointer, as \bhive{} read-protects the
unrolled microkernel's instructions page.

\section{Static extraction of memory-carried dependencies}\label{sec:staticdeps}

As put forwards by \anica~\cite{anica}, memory-carried dependencies are a
difficult challenge for code analyzers. This is further confirmed by the
results of \cesasme{}, presented in Section~\ref{ssec:memlatbound}.

To this end, we present \staticdeps{}, a heuristic-based static assembly
analyzer able to extract most of the memory-carried dependencies that are
relevant from a latency analysis perspective. It operates at the basic-block
level, and tracks dependencies across arbitrarily complex pointer arithmetic,
through arbitrarily many loop iterations.

Working at the basic-block level, it is however lacking context, as we argued
earlier; as such, it cannot detect aliasing stemming from outside of the loop.

\subsection{Relevance of memory dependencies in hardware}\label{ssec:rob}

Modern Out-of-Order (OoO) hardware has dedicated significant architectural
resources to maximize the pipeline usage. One such key component is the reorder
buffer (ROB) ---~an extension of Tomasulo's algorithm~\cite{tomasulo}~---, a
circular buffer storing incoming \uops{}, decoded and renamed from
instructions, in the frontend. When a \uop{} is committed, as soon as it is no
longer busy, the ROB can safely move its head pointer, freeing space for new
\uops{} at its tail.

In particular, if a \uop{} $\mu_1$ is not yet committed, the
ROB may not contain \uops{} more than the ROB's size ahead of $\mu_1$. This is
also true for instructions, as the vast majority of instructions
decode to at least one \uop{}\footnote{Some \texttt{mov} instructions from
register to register may, for instance, only have an impact on the renamer;
no \uops{} are dispatched to the backend.}.

Nevertheless, instruction level parallelism (ILP) might still be limited,
either by dependencies or other factors. Speculative techniques such as memory
disambiguation address that point, by for instance trying to detect whether
a given load depends on an earlier not-yet-committed store, in order to allow
safe out-of-order execution of non-dependent memory accesses. If such
speculation turns out wrong, execution is typically rolled back to before
the offending load in the ROB, incurring at least the penalty of another load.

Recent microarchitectures still lack ways of dealing with true data
dependencies, especially at the memory level, even if ongoing research with
value prediction~\cite{valueprediction-perais,valueprediction-sheikh} tackles
that issue, but to the best of our knowledge value prediction has not yet been
implemented in products of mainstream CPU vendors.

\subsection{Detecting different types of dependencies}

Depending on their type, some dependencies are significantly harder to
statically detect than others.

\textbf{Register-carried dependencies in straight-line code} are
easiest to detect, by keeping track of which instruction last wrote each
register in a \emph{shadow register file}. This is most often supported by code analyzers
---~for instance, \llvmmca{} and \uica{} support it.

\textbf{Register-carried, loop-carried dependencies}
can, to some extent, be detected the same way. As the basic block
is always assumed to be the body of an infinite loop, a straight-line analysis
can be performed on a duplicated kernel. This strategy is \eg{} adopted by
\osaca{}~\cite{osaca2} (§II.D).
When dealing only with register accesses, this
strategy is always sufficient: as each iteration always executes the same basic
block, it is not possible for an instruction to depend on another instruction
two iterations earlier or more.

\smallskip

\textbf{Memory-carried dependencies}, however, are significantly harder to tackle. While basic
heuristics can handle some simple cases, in the general case two main
difficulties arise:
\begin{enumerate}[label=(\roman*)]
    \item{}\label{memcarried_difficulty_alias} pointers may \emph{alias}, \ie{}
        point to the same address or array; for instance, if \reg{rax} points
        to an array, it may be that \reg{rbx} points to $\reg{rax} + 8$, making
        the detection of such a dependency difficult;
    \item{}\label{memcarried_difficulty_arith} arbitrary arithmetic operations
        may be performed on pointers, possibly through diverting paths: \eg{}
        it might be necessary to detect that $\reg{rax} + 16 << 2$ is identical
        to $\reg{rax} + 128 / 2$; this requires semantics for assembly
        instructions and tracking formal expressions across register values
        ---~and possibly even memory.
\end{enumerate}

Tracking memory-carried dependencies is, to the best of our knowledge, not done
in code analyzers, or only with simple heuristics, as our results with
\cesasme{} suggests.

\textbf{For loop-carried, memory-carried dependencies}, the strategy
previously used for register-carried dependencies is not sufficient
at all times when the dependencies tracked are memory-carried, as dependencies
may reach further back than the previous iteration, \eg{} in a dynamic
implementation of the Fibonacci sequence.

\subsection{The \staticdeps{} heuristic}

To statically detect memory dependencies, we use a randomized heuristic approach.
We consider the set $\calR$ of values
representable by a 64-bits unsigned integer; we extend this set to $\calRbar =
\calR \cup \{\bot\}$, where $\bot$ denotes an invalid value. We then proceed by
keeping track of which instruction last wrote each memory address written to,
using the following principles.
\begin{itemize}
    \item{} Whenever an unknown value is read, either from a register or from
        memory, generate a fresh value from $\calR$, uniformly sampled at
        random. This value is saved to a shadow register file or memory, and
        will be used again the next time this same data is accessed. Note that a 
        shadow is a data structure that describes the original component with a 
        given granularity and additional meta-information. For instance, 
        a shadow memory in our case is a set of memory locations, 
        each of which is associated with a value from $\calRbar$ and the last 
        instruction that wrote to it.

    \item{} Whenever an integer arithmetic operation is encountered, compute
        the result of the operation and save the result to the shadow register
        file or memory.

    \item{} Whenever another kind of operation or an unsupported operation
        is encountered, save the destination operand as $\bot$;
        this operation is assumed to not be valid pointer arithmetic.
        Operations on $\bot$ always yield $\bot$ as a result.

    \item{} Whenever writing to a memory location, compute the written address
        using the above principles, and keep track of the instruction that last
        wrote to a memory address.

    \item{} Whenever reading from a memory location, compute the read address
        using the above principles, and generate a dependency from the current
        instruction to the instruction that last wrote to this address (if
        known).
\end{itemize}

To handle loop-carried dependencies, we remark that
while far-reaching dependencies may \emph{exist}, they are not
necessarily \emph{relevant} from a performance analysis point of view. Indeed,
if an instruction $i_2$ depends on a result previously produced by an
instruction $i_1$, this dependency is only relevant if it is possible that
$i_1$ is not yet completed when $i_2$ is considered for issuing ---~else, the
result is already produced, and $i_2$ needs not wait to execute.
Given the structure of the ROB as detailed in Section~\ref{ssec:rob} above, it
seems that a dependency is relevant only if it does not reach more than the
size of the ROB away; we formalize and prove this intuition
in Section~\ref{ssec:rob_proof} below.

A possible solution to detect loop-carried dependencies in a kernel $\kerK$ is
thus to unroll it until it contains at least
$\card{\text{ROB}} + \card{\kerK}$ instructions.
This ensures that every instruction in the last kernel can find dependencies
reaching up to $\card{\text{ROB}}$ back.

On Intel CPUs, the reorder buffer size contained 224 \uops{} on Skylake (2015),
or 512 \uops{} on Golden Cove (2021)~\cite{wikichip_intel_rob_size}. These
sizes are small enough to reasonably use this solution without excessive
slowdown.

\subsection{Practical implementation}

We implement \staticdeps{} in Python, using \texttt{pyelftools} and the
\texttt{capstone} disassembler
to extract and disassemble the targeted basic
block. The semantics needed to compute encountered operations are obtained by
lifting the kernel's assembly to \valgrind{}'s \vex{} intermediary
representation.

The implementation of the heuristic detailed above provides us with a list
of dependencies across iterations of the considered basic block. We then
``re-roll'' the unrolled kernel: each dependency is written as a triplet
$(\texttt{source}, \texttt{dest}, \Delta{}k)$, where the first two
elements are the source and destination instruction of the dependency \emph{in
the original, non-unrolled kernel}, and $\Delta{}k$ is the number of iterations
of the kernel between the source and destination of the dependency.

Finally, we filter out spurious dependencies: each dependency found should
occur for each kernel iteration $i$ at which $i + \Delta{}k$ is within bounds.
If the dependency is found for less than $80\,\%$ of those iterations, the
dependency is declared spurious and is dropped.

\subsection{Long distance dependences can be ignored}\label{ssec:rob_proof}

\begin{definition}[Distance between instructions]
  Let $\left(I_p\right)_{0\leq p<n}$ be the trace of executed instructions.
  For $p<q$, $\distance{I_p}{I_{p'}}$ is the overall number of decoded \uops{}
  for the subtrace $\left(I_r\right)_{p\leq r\leq p'}$ minus one.
\end{definition}

\begin{thm}[Long distance dependencies]\label{thm.longdist}
A dependency between two instructions that are separated by at least $R$ others \uops{} can be ignored.
\end{thm}

To prove this assertion we need a few postulates that describe the functioning of a CPU and in particular how \uops{} transit in (decoded) and out (retired) the reorder buffer.
\begin{postulate}[Reorder buffer as a circular buffer]
  Reorder buffer is a circular buffer of size say $R$.
  It contains only decoded \uops{}.
Let us denote $i_d$ the \uop{} at position $d$ in the reorder buffer.
Assume $i_d$ just got decoded.
We have that for every $q$ and $q'$ in $[0,R)$:
    $$(q-d-1) \% R<(q'-d-1) \% R\ \Leftrightarrow \ i_q \textrm{ is decoded before } i_{q'}$$
\end{postulate}

  If a \uop{} has not been retired yet (issued and executed), it cannot be replaced by any freshly decoded instruction.
  In other words every non-retired decoded \uop{} are in the reorder buffer.
  This is possible thanks to the notion of \emph{full reorder buffer}:
\begin{postulate}[Full reorder buffer]
  Let us denote by $i_d$ the \uop{} that just got decoded.
  The reorder buffer is said to be full if for $q=(d+1) \% R$, \uop{} $i_q$ is not retired yet.
  If the reorder buffer is full, then instruction decoding is stalled.
\end{postulate}

  Let $(I_p)_{0\le p<n}$ be a trace of executed instructions.
  Each of these instructions are iteratively decoded, issued, and retired.
  We will also denote by $(i_q)_{0\le q<m}$ the trace of decoded \uops{}.
  To prove Theorem~\ref{thm.longdist} we need to state that two in-flight \uops{} are distant of at most $R$ \uops{}.
  For any instruction $I_p$, we denote as $Q_p$ the range of indices such that $(i_q)_{q\in Q_p}$ are the \uops{} from the decoding of $I_p$.
  We will call an \emph{in-flight} \uop{}, a \uop{} that is decoded but not yet retired;
  It is necessarily in the reorder buffer.

\begin{lemma}[Distance of in-flight \uops{}]
  For any pair of instructions $(I_p,I_{p'})$, and two corresponding \uops{}, $(i_q,i_{q'})$  with $(q,q')\in Q_p\times Q_{p'}$,
  $$\textrm{inflight}(i_q) \wedge \textrm{inflight}(i_{q'}) \Rightarrow \distance{I_p}{I_{p'}}<R$$
\end{lemma}

  In case of branch misprediction, some additional instructions might also be decoded (and potentially issued).
  In other words, $(i_q)_{0\le q<m}$ potentially contains more \uops{} than those matching the executed instruction.
  However, if not surjective, the relation from $\{I_p\}_{0\le p<n}$ to $\{i_q\}_{0\le q<m}$ is clearly injective and increasing (using the order within which instructions and \uops{} are listed).
  In other words, $\distance{I_p}{I_{p'}}\le |q'-q|$.
  Observe that at any time, the content of the ROB can be seen as a window of length $R$ over $(i_q)_{0\le q<m}$.
  Consequently, if both $i_q$  and $i_{q'}$ are in-flight then
  $|q'-q|<R$.~\hfill\qedsymbol

\begin{postulate}[Issue delay]
Reasons why the issue of a \uop{} $i$ is delayed can be:
\begin{enumerate}[noitemsep,topsep=0pt,parsep=0pt,partopsep=0pt]
  \item $i$ is not yet in the reorder buffer
  \item $i$ depends on \uop{} $i'$ which is not retired yet
  \item ports on which $i$ can be mapped are all occupied
\end{enumerate}
\end{postulate}

Theorem~\ref{thm.longdist} is now a direct consequence (proof by contradiction) of the previous observations.
Let us consider a delayed issue for \uop{} $i$ where the unique cause is a dependence from \uop{} $i'$, that is:
\begin{enumerate}[noitemsep,topsep=0pt,parsep=0pt,partopsep=0pt]
  \item $i$ is already in the reorder buffer
  \item $i$ depends on \uop{} $i'$ which is not retired yet
  \item at least one port on which $i$ can be mapped is available
\end{enumerate}
Since $i'$ is not retired yet and $i'$ is ``before'' $i$, $i'$ is still in the
reorder buffer, \ie{} both $i$ and $i'$ are in the reorder buffer.
\hfill\qedsymbol

\subsection{Limitations}\label{ssec:staticdeps_limits}

In its current implementation, \staticdeps{} is limited to read-after-write
dependencies, which we believe to be sufficient for latency analysis on usual
CPUs. However, our heuristic may easily support other types of dependencies
(\eg{} write-after-write) by tracking additional data if necessary.

\smallskip{}

We argued earlier that one of the shortcomings that most
crippled state-of-the-art tools was that analyses were conducted
out-of-context, considering only the basic block at hand. This analysis is also
true for \staticdeps{}, as it is still focused on a single basic block in
isolation; in particular, any aliasing that stems from outside the analyzed
basic block is not visible to \staticdeps{}.

Work towards a broader analysis range, \eg{} at the scale of a function, or at
least initializing values with gathered assertions ---~maybe based on abstract
interpretation techniques~--- could be beneficial to the quality of
dependencies detections.

\smallskip{}

As \staticdeps{}'s heuristic is based on randomness, it
may yield false positives: two registers could theoretically be assigned the
same value sampled at random, making them aliasing addresses. This is, however,
very improbable, as values are sampled from a set of cardinality $2^{64}$. If
necessary, the error can be reduced by amplification: running multiple times
the algorithm on different randomness seeds reduces the error exponentially.

Conversely, \staticdeps{} should not present false negatives due to randomness.
Dependencies may go undetected, \eg{} because of out-of-scope aliasing or
unsupported operations. However, no dependency that falls into the scope 
of our baseline \depsim{} ---~as briefly described in 
Section~\ref{ssec:staticdeps_eval_depsim}~--- should be missed because of random 
initialisations.

\smallskip{}

In silicon, dependencies are implemented not between instructions, but between
\uops{}. Modeling the decomposition of instructions to \uops{}, together with
providing semantics for each \uop{}, however, is not done to our knowledge.
Furthermore, it would need to be done for each microarchitecture: while
instructions' semantics are defined by the ISA, each microarchitecture may
split instructions to \uops{} in its own way.

\section{Experiments and evaluation}\label{sec:exp_setup}

Our experiments are twofold. We first analyze the results from \cesasme{}, from
which we confirm \anica{}'s case study: memory-carried dependencies are a
crucial shortcoming of state-of-the-art analyzers. We then evaluate
\staticdeps{}' capacity to detect these dependencies: we first compare the
dependencies detected to those actually present, then evaluate the gain in
precision when enriching \uica{}'s model with \staticdeps{}.

\subsection{Experimental environment}

The experiments presented in this paper were all realized on a Dell PowerEdge
C6420 machine from the Grid5000 cluster~\cite{grid5000}, equipped with 192\,GB
of DDR4 SDRAM ---~only a small fraction of which was used~--- and two Intel
Xeon Gold 6130 CPUs (x86-64, Skylake microarchitecture) with 16 cores each.

The experiments themselves were run inside a Docker environment very close to
our artifact, based on Debian Bullseye. Care was taken to disable
hyperthreading to improve measurements stability. For tools whose output is
based on a direct measurement (\perf, \bhive), the benchmarks were run
sequentially on a single core with no experiments on the other cores. No such
care was taken for \gus{} as, although based on a dynamic run, its prediction
is purely function of recorded program events and not of program measures. All
other tools were run in parallel.

We use \llvmmca{} \texttt{v13.0.1}, \iaca{} \texttt{v3.0-28-g1ba2cbb},
\gus{} at commit \texttt{87463c9},
\bhive{} at commit \texttt{5f1d500},
\uica{} at commit \texttt{9cbbe93},
\ithemal{} at commit \texttt{b3c39a8}.

\subsection{\cesasme{}: analyzers' throughput predictions}\label{ssec:overall_results}

Running \cesasme{}'s harness provides us with 3500
benchmarks ---~after filtering out non-L1-resident
benchmarks~---, on which each throughput predictor is run. We make the full
output of our tool available in our artifact.

The raw complete output from \cesasme{} ---~roughly speaking, a
large table with, for each benchmark, a cycle measurement, cycle count for each
throughput analyzer, the resulting relative error, and a synthesis of the
bottlenecks reported by each tool~--- enables many analyses that, we believe,
could be useful both to throughput analysis tool developers and users. Tool
designers can draw insights on their tool's best strengths and weaknesses, and
work towards improving them with a clearer vision. Users can gain a better
understanding of which tool is more suited for each situation.

\begin{table*}
    \centering
    \begin{tabular}{l r r r r r r r r r}
        \toprule
\textbf{Bencher} & \textbf{Datapoints} & \textbf{Failures} & \textbf{(\%)} &
\textbf{MAPE} & \textbf{Median} & \textbf{Q1} & \textbf{Q3} & \textbf{K. tau} & \textbf{Time (CPU$\cdot$h)}\\
\midrule
BHive & 2198 & 1302 & (37.20\,\%) & 27.95\,\% & 7.78\,\% & 3.01\,\% & 23.01\,\% & 0.81 & 1.37\\
llvm-mca & 3500 & 0 & (0.00\,\%) & 36.71\,\% & 27.80\,\% & 12.92\,\% & 59.80\,\% & 0.57 & 0.96 \\
UiCA & 3500 & 0 & (0.00\,\%) & 29.59\,\% & 18.26\,\% & 7.11\,\% & 52.99\,\% & 0.58 & 2.12 \\
Ithemal & 3500 & 0 & (0.00\,\%) & 57.04\,\% & 48.70\,\% & 22.92\,\% & 75.69\,\% & 0.39 & 0.38 \\
Iaca & 3500 & 0 & (0.00\,\%) & 30.23\,\% & 18.51\,\% & 7.13\,\% & 57.18\,\% & 0.59 & 1.31 \\
Gus & 3500 & 0 & (0.00\,\%) & 20.37\,\% & 15.01\,\% & 7.82\,\% & 30.59\,\% & 0.82 & 188.04 \\
\bottomrule
    \end{tabular}
    \caption{Statistical analysis of overall results}\label{table:overall_analysis_stats}
    \vspace{-1.5em}
\end{table*}

\begin{figure}
    \includegraphics[width=\linewidth]{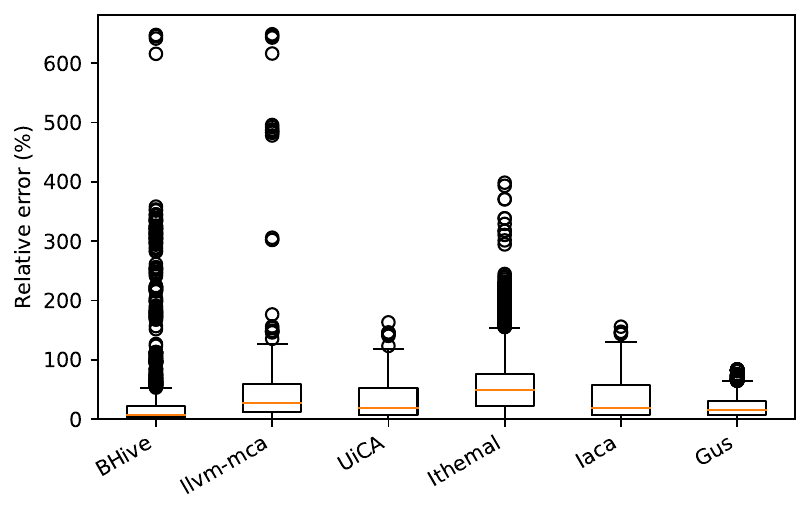}
    \vspace{-0.7cm}
    \caption{Statistical distribution of relative errors}\label{fig:overall_analysis_boxplot}
\end{figure}

The error distribution of the relative errors, for each tool, is presented as a
box plot in Figure~\ref{fig:overall_analysis_boxplot}. Statistical indicators
are also given in Table~\ref{table:overall_analysis_stats}. We also give, for
each tool, its Kendall's tau indicator~\cite{kendall1938tau}: this indicator,
used to evaluate \eg{} uiCA~\cite{uica} and Palmed~\cite{palmed}, measures how
well the pair-wise ordering of benchmarks is preserved, $-1$ being a full
anti-correlation and $1$ a full correlation. This is especially useful when one
is not interested in a program's absolute throughput, but rather in comparing
which program has a better throughput.

These results are, overall, significantly worse than what each tool's article
presents. We attribute this difference mostly to the specificities of
Polybench: composed of computation kernels, it intrinsically stresses the
CPU more than basic blocks extracted out of the Spec benchmark suite. This
difference is clearly reflected in the experimental section of the Palmed
article~\cite{palmed}: the accuracy of most tools is worse on Polybench than on
Spec, often by more than a factor of two.

As \bhive{} and \ithemal{} do not support control flow instructions
(\eg{} \texttt{jump} instructions), those had
to be removed from the blocks before analysis. While none of these tools, apart
from \gus{} ---~which is dynamic~---, is able to account for branching costs,
these two analyzers were also unable to account for the front- and backend cost
of the control flow instructions themselves as well ---~corresponding to the
$TP_U$ mode introduced by \uica~\cite{uica}, while others
measure $TP_L$.

\subsection{\cesasme{}: impact of dependency-boundness}\label{ssec:memlatbound}

An overview of the full results table (available in our artifact) suggests that 
on a significant number of rows, the static tools
---~thus leaving \gus{} and \bhive{} apart~---, excepted \ithemal, often yield
comparatively bad throughput predictions \emph{together}.
To confirm this observation, we look at the 30\,\% worst benchmarks ---~in
terms of MAPE relative to \perf~--- for \llvmmca, \uica{} and \iaca{}
---~yielding 1050 rows each. All of these share 869 rows (82.8\,\%), which we
call \textit{jointly bad rows}.
In the overwhelming majority (97.5\,\%) of those jointly bad rows, the tools predicted
fewer cycles than measured, meaning that a bottleneck is either missed or
underestimated.

\begin{figure}
    \includegraphics[width=\linewidth]{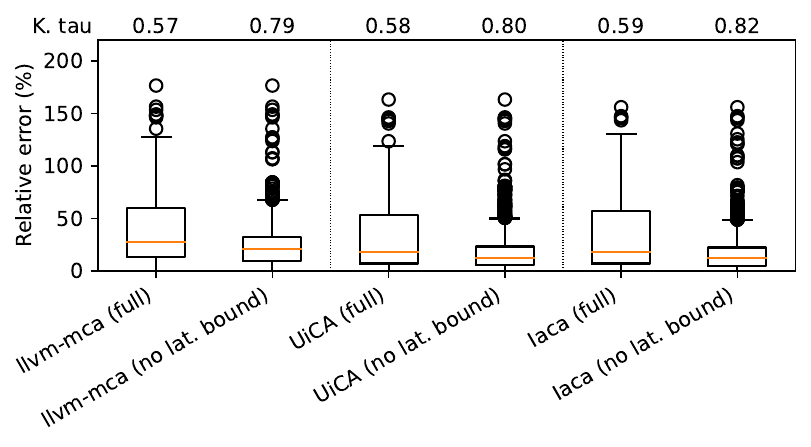}
    \vspace{-0.8cm}
    \caption{Statistical distribution of relative errors, with and without
    pruning latency bound through memory-carried dependencies rows (\llvmmca{}
outliers trimmed)}\label{fig:nomemdeps_boxplot}
\end{figure}

There is no easy way, however, to know for certain which of the 3500 benchmarks
are latency bound: no hardware counter reports this. We investigate this
further using \gus's sensitivity analysis: in complement of the ``normal''
throughput estimation of \gus, we run it a second time, disabling the
accounting for latency through memory dependencies. By construction, this
second measurement should be either very close to the first one, or
significantly below. We then assume a benchmark to be latency bound due to
memory-carried dependencies when it is at least 40\,\% faster when this latency
is disabled; there are 1112 (31.8\,\%) such benchmarks.

Of the 869 jointly bad rows, 745 (85.7\,\%) are declared latency
bound through memory-carried dependencies by \gus.
In Section~\ref{ssec:overall_results}, we presented in
Figure~\ref{fig:overall_analysis_boxplot} general statistics on the tools
on the full set of benchmarks. We now remove the 1112 benchmarks
flagged as latency bound through memory-carried dependencies by \gus{} from the
dataset, and present in Figure~\ref{fig:nomemdeps_boxplot} a comparative
box plot for the tools under scrutiny, with Kendall's tau coefficient.
While the results for \llvmmca, \uica{} and \iaca{} globally improved
significantly, the most noticeable improvements are the reduced spread of the
results and the Kendall's $\tau$ correlation coefficient's increase.

From this, we argue that detecting memory-carried dependencies is a weak point
in current state-of-the-art static analyzers, and that their results could be
significantly more accurate if improvements are made in this direction. We thus
confirm \anica{}'s case study on a large dataset of benchmarks, stemming from
real-world code.

\subsection{\Staticdeps{}: dependency coverage \wrt{} \depsim}\label{ssec:staticdeps_eval_depsim}

In order to assess \staticdeps{}' results, we implemented \depsim{}, a simple
dynamic analyzer instrumenting binaries using \valgrind{}. It analyzes and
reports memory dependencies of a given binary at runtime.

We use the binaries produced by \cesasme{} as a dataset, as we already assessed
its relevance and contains enough benchmarks to be statistically meaningful. We
also already have tooling and basic-block segmentation available for those
benchmarks, making the analysis more convenient.

For each binary previously generated by \cesasme{}, we use its cached basic
block splitting and occurrence count. Among each binary, we discard any basic
block with fewer than 10\,\% of the occurrence count of the most-hit basic
block as irrelevant.

For each of the considered binaries, we run our baseline dynamic analysis,
\depsim{}, and record its results.

For each of the considered basic blocks, we run \staticdeps{}. We translate the
detected dependencies back to original ELF addresses, and discard the
$\Delta{}k$ parameter, as our dynamic analysis does not report an equivalent
parameter, but only a pair of program counters. Each of the dependencies
reported by \depsim{} whose source and destination addresses belong to the
basic block considered are then classified as either detected or missed by
\staticdeps{}. Dynamically detected dependencies spanning across basic blocks
are discarded, as \staticdeps{} cannot detect them by construction.

We consider two metrics: the unweighted dependencies coverage, $\cov_u$, as
well as the weighted dependencies coverage, $\cov_w$:

\noindent\begin{minipage}{0.45\linewidth}
\[
    \cov_u = \dfrac{\card{\text{found}}}{\card{\text{found}} + \card{\text{missed}}}
\]
\end{minipage}\hfill\begin{minipage}{0.45\linewidth}
\[
    \cov_w = 
        \dfrac{
            \sum_{d \in \text{found}} \rho_d
        }{\sum_{d \in \text{found}\,\cup\,\text{missed}} \rho_d}
\]
\end{minipage}
\smallskip

where $\rho_d$ is the number of occurrences of the dependency $d$, dynamically
detected by \depsim.

\begin{table}
    \centering
    \begin{tabular}{r r r}
        \toprule
        \textbf{Lifetime} & $\cov_u$ (\%) & $\cov_w$ (\%) \\
        \midrule
        $\infty$ & 38.1\,\% & 44.0\,\% \\
        1024 & 57.6\,\% & 58.2\,\% \\
        512 & 56.4\,\% & 63.5\,\% \\
        \bottomrule
    \end{tabular}
    \caption{Unweighted and weighted coverage of \staticdeps{} on \cesasme{}'s
    binaries}\label{table:cov_staticdeps}
\end{table}

\begin{table*}[t!]
    \centering
    \footnotesize
    \begin{tabular}{l l r r r r r r r}
        \toprule
        \textbf{Dataset} & \textbf{Bencher} & \textbf{Datapoints} & \textbf{MAPE} & \textbf{Median} & \textbf{Q1} & \textbf{Q3} & \textbf{$K_\tau$}\\
\midrule
        \multirow{2}{*}{Full} & \uica{} & 3500 & 29.59\,\% & 18.26\,\% & 7.11\,\% & 52.99\,\% & 0.58\\
                              & + \staticdeps{} & 3500 & 19.15\,\% & 14.44\,\% & 5.86\,\% & 23.96\,\% & 0.81\\
\midrule
        \multirow{2}{*}{Pruned} & \uica{} & 2388 & 18.42\,\% & 11.96\,\% & 5.42\,\% & 23.32\,\% & 0.80\\
                                & + \staticdeps{} & 2388 & 18.77\,\% & 12.18\,\% & 5.31\,\% & 23.55\,\% & 0.80\\
\bottomrule
    \end{tabular}
    \caption{Evaluation through \cesasme{} of the integration of \staticdeps{}
    to \uica{}}\label{table:staticdeps_uica_cesasme}
\end{table*}

These metrics are presented for the 3\,500 binaries of \cesasme{} in the first
data row of \autoref{table:cov_staticdeps}. The obtained coverage, of about
40\,\%, is lower than expected.

However, as we already pointed, dependencies matter only if the source and
destination are not too far away. To this end, we introduce in \depsim{} a notion of \emph{dependency lifetime}.
As we do not have access without a heavy runtime slowdown to elapsed cycles in
\valgrind{}, we define a \emph{timestamp} as the number of instructions
executed since beginning of the program's execution; we increment this count at
each branch instruction to avoid excessive instrumentation slowdown.
We re-run the previous experiments with lifetimes of respectively 1\,024 and
512 instructions, which roughly corresponds to the order of magnitude of the
size of a reorder buffer; results can also be found in
\autoref{table:cov_staticdeps}. While the introduction of a 1\,024 instructions
lifetime greatly improves the coverage rates, both unweighted and weighted,
further reducing this lifetime to 512 does not yield significant enhancements.

\bigskip{}

The final coverage results, with a rough 60\,\% detection rate, are reasonable
and detect a significant proportion of dependencies; however, many are still
not detected.

This may be explained by the limitations studied in
\autoref{ssec:staticdeps_limits} above, and especially the inability of
\staticdeps{} to detect dependencies through aliasing pointers. This falls,
more broadly, into the problem of lack of context that we expressed before: we
expect that an analysis at the scale of the whole program, that would be able
to integrate constraints stemming from outside the loop body, would capture
many more dependencies.

\subsection{\Staticdeps{}: enriching \uica{}'s model}

\begin{figure}
    \centering
    \includegraphics[width=0.8\linewidth]{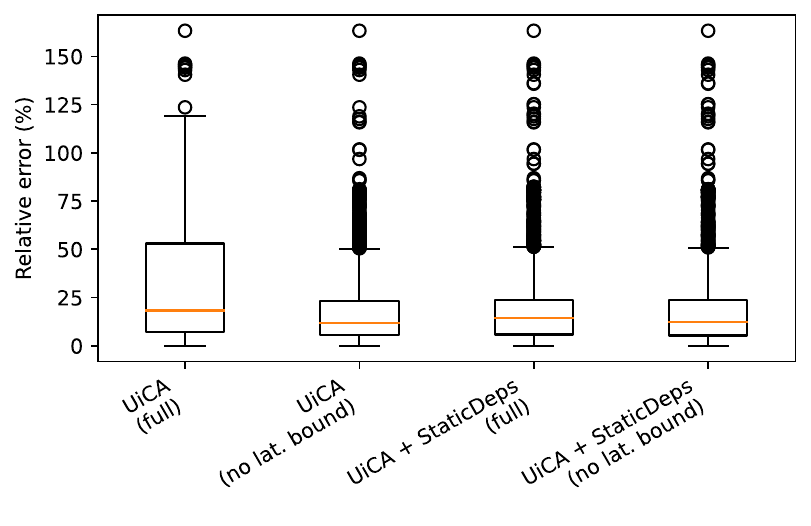}
    \vspace{-0.5cm}
    \caption{Statistical distribution of relative errors of \uica{}, with and
    without \staticdeps{} hints, with and without pruning latency bound through
memory-carried dependencies rows}\label{fig:staticdeps_uica_cesasme_boxplot}
\end{figure}

To estimate the real gain in performance debugging scenarios, however, we
integrate \staticdeps{} into \uica{}.

There is, however, a discrepancy between the two tools: while \staticdeps{}
works at the assembly instruction level, \uica{} works at the \uop{} level. In
real hardware, dependencies indeed occur between \uops{}; however, we are not
aware of the existence of a \uop{}-level semantic description of the x86-64
ISA, which made this level of detail unsuitable for the \staticdeps{} analysis.

We bridge this gap in a conservative way: whenever two instructions $i_1, i_2$
are found to be dependant, we add a dependency between each couple $\mu_1 \in
i_1, \mu_2 \in i_2$. This approximation is thus pessimistic, and should predict
execution times biased towards a slower computation kernel. A finer model, or a
finer (conservative) filtering of which \uops{} must be considered dependent
---~\eg{} a memory dependency can only come from a memory-related \uop{}~---
may enhance the accuracy of our integration.

We then evaluate our gains by running \cesasme{}'s harness by running both \uica{} and \uicadeps{}, as we did in
Section~\ref{ssec:memlatbound}, on two datasets:
first, the full set of 3\,500 binaries from \cesasme{}; then, the
set of binaries pruned to exclude benchmarks heavily relying on memory-carried
dependencies introduced in \autoref{ssec:memlatbound}. If \staticdeps{} is
beneficial to \uica{}, we expect \uicadeps{} to yield significantly better
results than \uica{} alone on the first dataset. On the second dataset,
however, \staticdeps{} should provide no significant contribution, as the
dataset was pruned to not exhibit significant memory-carried latency-boundness.
We present these results in \autoref{table:staticdeps_uica_cesasme}, as well as
the corresponding box-plots in \autoref{fig:staticdeps_uica_cesasme_boxplot}.

The full dataset \uicadeps{} row is extremely close, on every metric, to the
pruned, \uica{}-only row. On this basis, we argue that \staticdeps{}' addition
to \uica{} is very conclusive: the hints provided by \staticdeps{} are
sufficient to make \uica{}'s results as good on the full dataset as they were
before on a dataset pruned of precisely the kind of dependencies we aim to
detect. Furthermore, \uica{} and \uicadeps{}' results on the pruned dataset are
extremely close: this further supports the accuracy of \staticdeps{}.

While the results obtained against \depsim{} in
\autoref{ssec:staticdeps_eval_depsim} above were reasonable, they were not
excellent either, and showed that many kinds of dependencies were still missed
by \staticdeps{}. However, our evaluation on \cesasme{} by enriching \uica{}
shows that, at least on the workload considered, the dependencies that actually
matter from a performance debugging point of view are properly found.

This, however, might not be true for other kinds of applications that would
require a dependencies analysis.

\section{Conclusion}

In this article, we have presented the new heuristic-based
analyzer, \staticdeps{}, for statically identifying
memory-carried dependencies of a binary basic block.
We have shown that it can be used to beneficially extend
state-of-the-art code analyzers, such as \uica{}.
In order to provide sound baselines for evaluating such extensions,
we also have introduced the evaluation framework \cesasme{}, which is meant to
compose measurement environments and code analyzers and make their results
commensurable.
In order to take into account the impact of a basic block's context on its
measured throughput, the method we propose consists in reasoning at kernel-level
by lifting the basic block-level predictions.
Finally, \cesasme{} has allowed us to exhibit a significant
gain in accuracy,
thanks to \staticdeps{}, on latency-bound microbenchmarks
generated from the HPC-native Polybench benchmark suite.

\clearpage
\bibliographystyle{ACM-Reference-Format}
%\citsetyle{acmnumeric}
\bibliography{main}

\end{document}